\documentclass[conference]{IEEEtran}
\IEEEoverridecommandlockouts

\usepackage{amsmath,amssymb,amsfonts,mathtools}

\usepackage{graphicx}
\usepackage{textcomp}
\usepackage{xcolor}

\usepackage{multirow}
\usepackage{enumitem}
\usepackage[noadjust]{cite}

\usepackage[ruled,vlined]{algorithm2e}
\usepackage{tcolorbox}
\tcbuselibrary{skins,breakable,theorems}

\usepackage[caption=false,font=footnotesize]{subfig}
\usepackage{url}

\makeatletter
\long\def\@makecaption#1#2{%
  \vskip\abovecaptionskip
  \sbox\@tempboxa{#1. #2}%
  \ifdim \wd\@tempboxa >\hsize
    \footnotesize #1. #2\par
  \else
    \hbox to\hsize{\hfil \footnotesize #1. #2\hfil}%
  \fi
  \vskip\belowcaptionskip}
\makeatother

\newcommand{\mat}[1]{\mathbf{#1}}
\def\BibTeX{{\rm B\kern-.05em{\sc i\kern-.025em b}\kern-.08em
    T\kern-.1667em\lower.7ex\hbox{E}\kern-.125emX}}

\newtcbtheorem[auto counter, number within=section]{femalgorithm}{Algorithm}
{enhanced,
  colback=white,
  colframe=black,
  fonttitle=\bfseries}{alg}
  
\begin{document}

\title{Robustness Verification of Binary Neural Networks: An Ising and Quantum-Inspired Framework}
\author{\IEEEauthorblockN{
Rahul Singh\textsuperscript{$\dagger$,1},
Seyran Saeedi\textsuperscript{2},
Zheng Zhang\textsuperscript{1}}
\IEEEauthorblockA{\textsuperscript{1}Department of Electrical and
Computer Engineering, University of California, Santa Barbara, CA, U.S.A.}
\IEEEauthorblockA{\textsuperscript{2}Independent}
\IEEEauthorblockA{\ \textsuperscript{$\dagger$}Corresponding author: rps@ucsb.edu}
}
\maketitle

\begin{abstract}
Binary neural networks (BNNs) are increasingly deployed in edge computing applications due to their low hardware complexity and high energy efficiency. However, verifying the robustness of BNNs against input perturbations, including adversarial attacks, remains computationally challenging because the underlying decision problem is inherently combinatorial. In this paper, we propose an Ising- and quantum-inspired framework for BNN robustness verification. We show that, for a broad class of BNN architectures, robustness verification can be formulated as a Quadratic Constrained Boolean Optimization (QCBO) problem and subsequently transformed into a Quadratic Unconstrained Boolean Optimization (QUBO) instance amenable to Ising and quantum-inspired solvers. We demonstrate the feasibility of this formulation on binarized MNIST by solving the resulting QUBOs with a free energy machine (FEM) solver and simulated annealing. We also show the deployment of this framework on quantum annealing and digital annealing platforms. Our results highlight the potential of quantum-inspired computing and Ising computing as a pathway toward trustworthy AI systems.
\end{abstract}


\section{Introduction}
Deep neural networks (DNNs) have achieved success across domains such as natural language processing, computer vision, and scientific computing, but they are vulnerable to small input perturbations~\cite{DBLP:journalscorrSzegedyZSBEGF13} that can lead to incorrect or dangerous predictions. This vulnerability is especially problematic in safety-critical applications, including autonomous driving, healthcare, and finance. Consequently, robustness verification—certifying correctness under bounded input perturbations—is essential for trustworthy AI systems.

In resource-constrained or edge environments, the computational and memory costs of full-precision DNNs can be prohibitive. Binary neural networks (BNNs), which restrict weights and activations to binary values, provide reduced hardware complexity, smaller memory footprints, and energy-efficient bitwise operations~\cite{courbariaux2016binaryconnecttrainingdeepneural,courbariaux2016binarizedneuralnetworkstraining,QIN2020107281, rastegari2016xnornetimagenetclassificationusing}. Although BNNs enable efficient deployment on FPGAs, ASICs, and microcontrollers, they remain vulnerable to adversarial and structured perturbations. Robustness verification of BNNs is therefore crucial for reliable edge deployment and for understanding the limitations of aggressively quantized models. Existing approaches primarily rely on classical exact solvers such as SAT, MILP, and SMT tools~\cite{narodytska2018verifying,katz2017reluplex}, but the resulting combinatorial optimization problems are intrinsically expensive on CPUs and GPUs, often exhibiting exponential worst-case complexity.

Recent advances in quantum-inspired computing hardware~\cite{preskill2018quantum,johnson2011quantum} have enabled practical systems such as digital annealers~\cite{Aramon2019}, Ising machines~\cite{yamamoto2017coherent,inagaki2016coherent}, and free-energy solvers~\cite{shen2025free}. These platforms natively support Ising or QUBO formulations that capture many NP-hard combinatorial optimization problems. Their potential advantage lies in massive parallelism, collective dynamics, and stochastic or coherent evolution for exploring nonconvex energy landscapes. For example, quantum annealers and coherent Ising machines can escape local minima in problems such as Max-Cut, SAT and graph partitioning, while quantum-inspired solvers have shown competitive performance in portfolio optimization~\cite{venturelli2019reverse}, traffic routing~\cite{neukart2017traffic}, and protein-folding approximations~\cite{perdomo2012finding}. Although large-scale fault-tolerant quantum advantage remains an open challenge, these platforms already offer practical benefits for specific structured optimization tasks.

In this paper, we propose an Ising and quantum-inspired approach for robustness verification of BNN. We formulate BNN robustness verification as a Quadratic Constrained Boolean Optimization (QCBO) problem and transform it into a Quadratic Unconstrained Binary Optimization (QUBO) representation compatible with quantum annealer and Ising machines. Our contributions are two-fold:
\begin{itemize}
\item \textbf{Theoretical and algorithmic formulation:}
We show that robustness verification of a generic $C$-class BNN can be reformulated as a Quadratic Unconstrained Boolean Optimization (QUBO) instance. Unlike MILP- or SAT-based encodings for classical solvers, our formulation directly matches the native input format of Ising and annealing-style hardware, enabling a unified interface with quantum annealing, gate-based quantum algorithms (e.g., QAOA), and digital or analog Ising machines.

\item \textbf{Empirical validation and solver benchmarking:}
We verify the robustness of a ten-class BNN trained on MNIST using simulated annealing and free energy machine (FEM)~\cite{Shen2025}. We further demonstrate the deployment of this framework on practical D-wave quantum annealer and commercial digital annealer. The results demonstrate that the proposed QUBO mapping is practically solvable for nontrivial BNNs and extensible to emerging quantum-inspired hardware.
\end{itemize}

To the best of our knowledge, this work presents the first quantum-inspired pipeline for robustness verification of BNNs, providing a foundation for scalable robustness analysis and future theoretical and practical extensions.

\section{Background}

This section presents an overview of BNNs, robustness verification and the QUBO framework.

\subsection{Binary Neural Network Architecture}
\label{Sec:BNN}
The BNN studied here is a classification model that uses the element-wise sign function (\texttt{sgn}) as its non-linear activation and an argmax operation (\texttt{argmax}) in the output layer to determine the predicted class. These functions are defined as:
\begin{equation}
    \begin{split}
        \texttt{sgn}(x) &= \frac{x}{|x|}, \quad \text{for } x \neq 0, \\
        \underset{x \in \mat{S}}{\texttt{argmax}} \left(f(x)\right) &= \{x \in \mat{S} \mid f(s) \leq f(x),\ \forall s \in \mat{S}\}. 
    \end{split}
\end{equation}

The BNN consists of $L$ feed-forward layers with binary input $\mat{x}$ and output label $y$. Layer transformations are defined recursively as:
\begin{equation}
    \begin{split}
        \mat{y}^1 &= \texttt{sgn}(\mat{W}^0 \mat{x}), \\
        \mat{y}^{l+1} &= \texttt{sgn}(\mat{W}^{l} \mat{y}^{l}), \quad \text{for } l \in \{1, \ldots, L - 1\}, \\
        y &= \texttt{argmax} \left( \mat{W}^L \mat{y}^L \right).
    \end{split}
\end{equation}

We consider a $C$-class BNN (e.g., $C=10$ or $2$) trained on MNIST, applying dimensionality reduction to the input and removing duplicate samples with identical inputs. Although the data set used contains unique input-output pairs, the formulation itself does not depend on the uniqueness of the data set.

\subsection{Neural Network Robustness Verification}
\label{SubSec:RobustnessVerification}
Given a neural network $F$, an input data sample $\mat{x} \in \mathcal{X}$, and its associated label $y = F(\mat{x})$, a robustness verification tool aims to determine whether any perturbation $\boldsymbol{\tau} \in \boldsymbol{\Omega}$, where $\boldsymbol{\Omega}$ denotes the set of allowed perturbations, causes a change in the classification output. In other words, the neural network is said to be \textit{robust} in $\boldsymbol{\Omega}$ if  
\begin{equation}
    F(\mat{x} + \boldsymbol{\tau}) = y, \quad \forall \; \boldsymbol{\tau} \in \boldsymbol{\Omega}.
\end{equation}

In contrast, the neural network is considered \textit{not robust} if there exists a perturbation $\boldsymbol{\tau} \in \boldsymbol{\Omega}$ such that $ F(\mat{x} + \boldsymbol{\tau}) \neq y$.

\subsection{Quadratic Unconstrained Boolean Optimization (QUBO)}
Recent advances~\cite{10.3389fphy.2014.00005} suggest that Ising solvers offer a promising alternative to solve NP-hard combinatorial optimization problems by reformulating them as Ising Hamiltonians. A widely studied and hardware compatible simplification of the Ising model is the Quadratic Unconstrained Binary Optimization (QUBO)~\cite{Glover2022}. For a Boolean input vector $\mat{x} \in \{0,1\}^n$, a QUBO is defined as:
\begin{equation}
    \text{QUBO}(\mat{x}) = \underset{\mat{x}}{\text{minimize}} \quad \mat{x}^\top \mat{Q} \mat{x}, \quad \mat{Q} \in \mathbb{R}^{n \times n}.
\end{equation}
The corresponding Ising form can be obtained by converting the Boolean variables $x_i \in \{0,1\}$ into spin variables $s_i \in \{-1, 1\}$ using the linear transformation:
\begin{equation}
    s_i = 2x_i - 1.
    \label{eq:BoolSpin}
\end{equation}

QUBO has emerged as a standard problem formulation in many unconventional computing platforms, including Ising machines, quantum annealers, and digital annealing hardware. For instance, \cite{Bian2017} demonstrated the application of a quantum annealer to solve QUBO problems, while \cite{Egger2021warmstartingquantum} solved QUBOs on gate based NISQ devices.

\section{Method}
\label{Sec:Method}

We consider a trained BNN, and search for a perturbation that induces misclassification for a given input. The verification task is reformulated as a minimization problem, where each binary operation is encoded as quadratic constraints over spin variables, yielding a quadratic constrained spin optimization. This formulation is then converted into a Boolean optimization problem, and finally all constraints are combined into a single QUBO Hamiltonian using a penalty construction.

\subsection{Minimization Problem}
\label{miniproblem}
The critical first step in our approach is to reformulate the robustness verification problem from Section~\ref{SubSec:RobustnessVerification} as an explicit \textit{minimization} problem. Let $\mat{x}$ denote the original (Boolean) input to the BNN, and let $\mat{x'}$ denote a perturbed input obtained by applying a binary perturbation mask $\boldsymbol{\tau}$:
\[
\mat{x'} = \mat{x} \oplus \boldsymbol{\tau}, \qquad \boldsymbol{\tau} \in \{0,1\}^d,
\]
where $\oplus$ denotes bitwise XOR and $d$ is the input dimension. Here, $\tau_i = 1$ indicates that pixel $i$ is flipped, while $\tau_i = 0$ means it is unchanged.

Our goal is to identify the minimal perturbation $\boldsymbol{\tau}$ that changes the BNN’s output, i.e., $F(\mat{x}) \neq F(\mat{x'})$, under a bounded perturbation budget. If such a perturbation exists, the BNN is declared \textit{non-robust} for $\mat{x}$; otherwise, it is robust within the specified bound. We quantify the change between $\mat{x}$ and $\mat{x'}$ using a simple {\it perceptual similarity measure} defined by the perturbation mask $\boldsymbol{\tau}$:
\begin{equation}
    \label{eq:BooleanPerceptualSimilarity}
    d(\mat{x}, \mat{x'}) = \sum_i \tau_i,
\end{equation}
which counts the number of perturbed pixels.

The robustness verification problem is formally written as
\begin{equation}
\label{eq:MiniProb}
\begin{split}
    \underset{\boldsymbol{\tau}}{\text{min}} \quad & d(\mat{x}, \mat{x'}) \\[0.25em]
    \text{s.t.} \quad & d(\mat{x}, \mat{x'}) \leq \epsilon, \quad
    F(\mat{x}) \neq F(\mat{x'}), \\[0.25em]
    & \mat{x'} = \mat{x} \oplus \boldsymbol{\tau}, \quad
     \boldsymbol{\tau} \in \{0,1\}^d.
\end{split}
\end{equation}
By casting verification as a search for minimal perceptual-change perturbations, we reduce it to a discrete combinatorial optimization problem.

\subsection{Quadratic Constrained Spin Optimization (QCSO)}
To verify a BNN on quantum-inspired hardware, we express both the model and verification objective as constraints on binary variables within a Mixed-Integer Linear Programming (MILP) framework~\cite{narodytska2018verifying}. This formulation focuses on two composite operations, \texttt{sgn} and \texttt{argmax}, which must be encoded as constraints over spin variables, while linear and multiplicative operations are handled using auxiliary variables.

We introduce auxiliary spin variables $z_{ij}^0$ to represent the product of the $j^{\text{th}}$ input and the weight of the $i^{\text{th}}$ neuron in the first layer. Similarly, $z_{ij}^l$ captures the corresponding product at the $l^{\text{th}}$ layer, for $l \in \{1, 2, \ldots, L\}$. The BNN is constructed using spin variables and integer labels, where $z_{ij}^l, w_{ij}^l, y^{l}i \in {-1, 1}$ and $y \in \mathbb{Z} \cap [0, C-1]$. The spin-based BNN from Section~\ref{Sec:BNN} is defined as:
\begin{equation}
    \label{eq:SpinBNN}
    \begin{split}
        z_{ij}^0 &= w_{ij}^0 x_j, \quad
        y^1_i = \texttt{sgn} \left(\sum_j z_{ij}^0\right), \\
        z_{ij}^l &= w_{ij}^l y_j^l, \quad
        y^{l+1}_i = \texttt{sgn}\left(\sum_j z_{ij}^l\right), \\
        y &= \texttt{argmax}\left(\sum_j z_{ij}^L\right),
    \end{split}
\end{equation}
where $x_j$ denotes the $j^{\text{th}}$ input, and $L$ represents the final layer.

Both \texttt{sgn} and \texttt{argmax} introduce non-linearities that are difficult to encode directly. The \texttt{sgn} function can be represented by enforcing non-negativity of the product between $\sum_j z_{ij}^l$ and $y^{l+1}i$:
\begin{equation}
    \label{eq:SignSim}
    y^{l+1}_i = \texttt{sgn}\left(\sum_j z_{ij}^l\right) \Leftrightarrow \left(\sum_j z_{ij}^l\right) y^{l+1}_i \geq 0.
\end{equation}
For \texttt{argmax}, we introduce auxiliary variables $r_i$:
\[
r_i = \texttt{sgn}\left(\sum_j \left(z_{ij}^L - z_{yj}^L\right)\right) \quad \forall i \neq y,
\]
where $y$ is the true label. This leads to the following constraint ensuring that the correct class (out of total $C$ classes) has the maximum activation:
\begin{equation}
    \label{eq:ArgmaxSimCorrect}
        y = \texttt{argmax}\left(\sum_j z_{ij}^L\right)
        \Leftrightarrow \sum_i r_i + C - 1 = 0.
\end{equation}
The resulting constraints for the spin-based BNN are summarized as:
\begin{equation}
    \label{eq:SpinBNNSim}
    \begin{split}
        &z_{ij}^0 = w_{ij}^0 x_j, \quad
        \left(\sum_j z_{ij}^0\right) y^1_i \geq 0, \\
        &z_{ij}^l = w_{ij}^l y_j^l, \quad
        \left(\sum_j z_{ij}^l\right) y^{l+1}_i \geq 0, \\
        &\left(\sum_j \left(z_{ij}^L - z_{yj}^L\right)\right) r_i \geq 0 \hspace{0.2cm} \forall \, i \neq y, \\
        &\sum_i r_i + C - 1 = 0.
    \end{split}
\end{equation}
For robustness verification, the objective is to find a perturbed input $\mat{x'}$ that causes misclassification. This is enforced by modifying the final constraint:
\begin{equation}
    \label{eq:ArgmaxSimIncorrect}
        F(\mat{x}) \neq F(\mat{x'}) 
        \Leftrightarrow \sum_i r_i + C - 1 > 0.
\end{equation}
To quantify perturbations, we define the {\it perceptual similarity for spin variables}, $d_s$, between $\mat{x}$ and its perturbed value $\mat{x'}$
\begin{equation}
    \label{eq:Perturbation}
    \begin{split}
        d_s(\mat{x}, \mat{x'}) &= \sum_i \frac{1 - s_{\tau_i}}{2}, \\
        x'_i &= x_i\, s_{\tau_i},
    \end{split}
\end{equation}
where $s_{\tau_i} \in \{-1, 1\}$ is a spin-valued perturbation variable indicating the perturbation applied to the $i$-th input bit.   

Finally, integrating \eqref{eq:MiniProb}, \eqref{eq:SpinBNNSim}, and \eqref{eq:Perturbation}, we obtain the QCSO formulation:
\begin{equation}
    \label{eq:SpinQCBO}
    \centering
    \begin{split}
    \min \quad & d_s(\mat{x}, \mat{x'}), \\[2ex]
    \text{s.t.} \quad & d_s(\mat{x}, \mat{x'}) \leq \epsilon, \\[2ex]
    & x'_i = x_i s_{\tau_i}, \quad z_{ij}^0 = w_{ij}^0 x'_j, \quad z_{ij}^l = w_{ij}^l y_j^l, \\[2ex]
    & \left(\sum_j z_{ij}^l\right) y^{l+1}_i \geq 0, \quad
    \left(\sum_j \left(z_{ij}^L - z_{kj}^L\right)\right) r_i \geq 0, \\[2ex]
    & \sum_i r_i + C - 1 > 0.
    \end{split}
\end{equation}
This formulation encodes the spin-based BNN and enables formal robustness verification via adversarial perturbation search under a perceptual similarity constraint.

\subsection{Quadratic Constrained Boolean Optimization}
\label{QCBO}

Quadratic constrained Boolean optimization (QCBO) is the Boolean equivalent of QCSO. All variables in QCBO are Boolean; spin variables from QCSO are converted to Boolean variables using~\eqref{eq:BoolSpin}. We denote the Boolean counterpart of any spin variable $s$ by $q_s$. The Boolean analogue of the spinwise product is the \texttt{XNOR} gate, which implements modulo-2 multiplication. Thus, the layerwise products in~\eqref{eq:SpinBNN} are replaced by \texttt{XNOR} operations between Boolean weight and activation variables:
\begin{equation}
    \label{eq:XNOR}
    \begin{split}
        q_{z_{ij}^0} &= \texttt{XNOR}(q_{w_{ij}^0}, q_{x_j}),\\
        q_{z_{ij}^l} &= \texttt{XNOR}(q_{w_{ij}^l}, q_{y_j^l}).
    \end{split}
\end{equation}

The nonlinear sign function acting on spin variables is replaced in QCBO by Boolean sign operators. We use two closely related but distinct Boolean sign functions:
\begin{itemize}
  \item $\texttt{sgn}_b(\cdot)$ operates on the \emph{integer sum} of Boolean variables and returns a Boolean output (see Section~\ref{Sec:Penalty} for its implementation). It is used inside each neuron to mimic the sign of the corresponding spin sum.
  \item $\texttt{sgn}_2(\cdot)$ acts on an integer represented in two's complement\footnote{In a two's-complement binary encoding, signed integers are represented so that the most significant bit indicates the sign and the remaining bits encode the magnitude modulo a power of two. For instance, in a 4-bit two's complement system, the binary sequence from $0000$ to $0111$ represent the integers $0$ to $+7$, while from $1000$ to $1111$ represent $-8$ to $-1$, with negative values obtained by inverting all bits of the corresponding positive value and adding one. } form and returns a Boolean flag indicating the sign of that integer.  It is used only in the output layer to encode the $\arg\max$ condition via the sign of the difference between class scores.
\end{itemize}

Operationally, $\texttt{sgn}_b$ is applied to sums of Boolean activations in all layers, where it reproduces the sign of the corresponding spin sums in the QCSO formulation. In contrast, $\texttt{sgn}_2$ is applied only in the output layer to the integer differences $\sum_j (q_{z_{ij}^L} - q_{z_{yj}^L})$, encoded in two's complement, and returns a single Boolean bit indicating whether a competing class score exceeds that of the true label. This mirrors the role of the auxiliary variables $r_i$ in the spin-based $\arg\max$ constraint and cleanly separates the neuron-wise nonlinearity from the final class comparison in the Boolean setting.

Using this notation, the Boolean formulation of the BNN layers becomes
\begin{equation}
    \label{eq:BoolBNN}
    \begin{split}
        &q_{z_{ij}^0} = \texttt{XNOR}(q_{w_{ij}^0}, q_{x_j}),\quad
        q_{y^{1}_i} = \texttt{sgn}_b\!\left(\sum_j  q_{z_{ij}^0}\right),\\
        &q_{z_{ij}^l} = \texttt{XNOR}(q_{w_{ij}^l}, q_{y_j^l}), \quad
        q_{y^{l+1}_i} = \texttt{sgn}_b\!\left(\sum_j  q_{z_{ij}^l}\right),\\
        &q_{r_i} = \texttt{sgn}_2\!\left(\sum_j \big(q_{z_{ij}^L} - q_{z_{yj}^L}\big)\right)\;
        \forall\, i \neq y,\quad \sum_i q_{r_i} = 0.
    \end{split}
\end{equation}
The perturbation in the input is simply added as $q_{x'_i} = q_{x_i} + \tau_i$. In Boolean form, this corresponds to an \texttt{XOR} between the binary input variable $q_{x_i}$ and the perturbation variable $\tau_i \in\{0,1\}$. Similar to QCSO, a constrained Boolean formulation (QCBO) is obtained by combining the BNN in~\eqref{eq:BoolBNN} with the output change constraint and the perturbation model:
\begin{equation}
    \label{eq:BoolQCBO}
    \begin{split}
        \min \ & d(q_{\mat{x}}, q_{\mat{x'}}) \\[0.2em]
        \text{s.t. } \ &
        d(q_{\mat{x}}, q_{\mat{x'}}) \leq \epsilon,\\
        & q_{x'_i} = \texttt{XOR}(q_{x_i}, \tau_i),\\
        & q_{z_{ij}^0} = \texttt{XNOR}(q_{w_{ij}^0}, q_{x'_j}),\quad
         q_{y^{1}_i} = \texttt{sgn}_b\!\left(\sum_j  q_{z_{ij}^0}\right),\\
         & q_{z_{ij}^l} = \texttt{XNOR}(q_{w_{ij}^l}, q_{y_j^l}),\quad
        q_{y^{l+1}_i} = \texttt{sgn}_b\!\left(\sum_j  q_{z_{ij}^l}\right),\\
         &q_{r_i} = \texttt{sgn}_2\!\left(\sum_j \big(q_{z_{ij}^L} - q_{z_{yj}^L}\big)\right)
          \; \forall i \neq \mat{y},\quad
         \sum_i q_{r_i} > 0.
    \end{split}
\end{equation}

\subsection{Constraint Simplification}

Boolean operators such as \texttt{XNOR} and \texttt{XOR} introduce higher complexity than simpler gates like \texttt{BUFFER} and \texttt{NOT}, mainly due to their more involved penalty functions (Table~\ref{Tab:PenaltyFunction}). In the spin-based BNN framework, substantial simplifications are possible because the weights are fixed (pre-trained) and the original input is known in robustness verification. As a result, \texttt{XNOR} operations in BNN layers can be reduced to \texttt{BUFFER} or \texttt{NOT}, significantly simplifying the constraints. For intermediate spin variables $q_{z_{ij}^l}$, which represents the \texttt{XNOR} of the weight bit $q_{w_{ij}^l}$ and neuron activation bit $q_{y_j^{l}}$, the simplification is

\begin{equation}
    \label{eq:XNORSim}
    \begin{split}
        &q_{z_{ij}^l} = \texttt{XNOR}(q_{w_{ij}^l}, q_{y_j^{l}}) \\
        &\implies 
        \begin{cases}
            q_{z_{ij}^l} = \texttt{NOT}(q_{y_j^{l}}) & \text{if } q_{w_{ij}^l} = 0, \\[3ex]
            q_{z_{ij}^l} = \texttt{BUFFER}(q_{y_j^{l}}) & \text{if } q_{w_{ij}^l} = 1.
        \end{cases}
    \end{split}
\end{equation}

In the first layer, perturbation and network constraints must be handled jointly. As described earlier, $q_{x'i}$ is the \texttt{XOR} of fixed input $q_{y_i^0}$ and perturbation $\tau_i$, followed by computing $q_{z_{ij}^0}$ via \texttt{XNOR} between fixed weight $q_{w_{ij}^0}$ and the perturbed bit $q_{x'_j}$. Combining \texttt{XOR} and \texttt{XNOR} yields

\begin{equation}
    \label{eq:XNORSimPerturb}
    \begin{split}
        &q_{x'_i} = \texttt{XOR}(q_{x_i}, \tau_i), \quad q_{z_{ij}^0} = \texttt{XNOR}(q_{w_{ij}^0}, q_{x'_j}) \\[3ex]
        &\implies 
        \begin{cases}
            q_{z_{ij}^0} = \texttt{NOT}(\tau_i) & \text{if } q_{w_{ij}^0} = 0, \, q_{x_j} = 0, \\[3ex]
            q_{z_{ij}^0} = \texttt{BUFFER}(\tau_i) & \text{if } q_{w_{ij}^0} = 1, \, q_{x_j} = 0, \\[3ex]
            q_{z_{ij}^0} = \texttt{BUFFER}(\tau_i) & \text{if } q_{w_{ij}^0} = 0, \, q_{x_j} = 1, \\[3ex]
            q_{z_{ij}^0} = \texttt{NOT}(\tau_i) & \text{if } q_{w_{ij}^0} = 1, \, q_{x_j} = 1.
        \end{cases}
    \end{split}
\end{equation}

This decomposition substantially reduces the complexity of the penalty terms, enabling more efficient optimization and verification.

\begin{table}[t]
\centering
\begin{tabular}{|c|c|c|}
\hline
\textbf{Gate} & \textbf{Operation} & \textbf{Penalty Function} \\ \hline
\texttt{XOR}   & $ q_k = q_i \oplus q_j $  & $ q_i + q_j - 2q_i q_j - q_k $ \\ \hline
\texttt{XNOR}  & $ q_k = q_i \odot q_j $   & $ 1 - q_i - q_j + 2q_i q_j - q_k $ \\ \hline
\texttt{BUFFER} & $ q_k = q_i $             & $ q_i + q_k - 2 q_i q_k $ \\ \hline
\texttt{NOT}   & $ q_k = \neg q_i $        & $ 2 q_i q_k - q_i - q_k $ \\ \hline
\end{tabular}
\caption{Penalty Functions for Boolean Operations}
\label{Tab:PenaltyFunction}
\end{table}

\subsection{Penalty Method: QCBO to QUBO}
\label{Sec:Penalty}

The penalty method transforms a constrained Boolean problem (QCBO) into an unconstrained QUBO by adding penalty terms for each constraint. Similar to the QCSO-to-Ising construction, each QCBO constraint is encoded as a quadratic penalty in the QUBO energy. This requires representing Boolean constraints as linear or quadratic polynomials, often introducing auxiliary variables for complex cases~\cite{10.3389fict.2017.00029}. We define local penalty functions for Boolean gates and sign layers and combine them into a global QUBO objective.
 
\subsubsection{Local Penalty Functions for Gates and Sign Layers}
\label{Sec:LocalPenalties}

\paragraph{Boolean gates}
Penalty functions for Boolean gates are derived from their truth tables. Given input and output variables, we construct a polynomial that evaluates to zero when the gate relation is satisfied and to a positive value otherwise. Thus, valid assignments incur zero penalty, while invalid ones are penalized. Penalty functions for \texttt{NOT} and \texttt{BUFFER} gates are listed in Table~\ref{Tab:PenaltyFunction}, with analogous constructions used for \texttt{XOR} and \texttt{XNOR} constraints in the BNN layers.

\paragraph{Sign layer}
We use two sign-function implementations: $\texttt{sgn}_b$ within neurons and $\texttt{sgn}_2$ for simplifying the output-layer $\arg\max$ constraint. The 
$\texttt{sgn}_b$ formulation requires that the number of binary variables being summed is 
$2^n - 1$ where $n \in \mathbb{N}$, ensuring symmetry around zero and allowing the sign to be represented by the most significant bit of the binary-encoded sum.

For example, with three binary spins \(x_1, x_2, x_3\) and output \(y\), the spin sum takes values in \(\{-3,-1,1,3\}\), whose signs map directly to the most significant bit of the corresponding binary encodings. This motivates using that bit as the output variable. The \(\texttt{sgn}_b\) with three inputs can be encoded via Boolean variables $q_{x_1}, q_{x_2}, q_{x_3}$ and an auxiliary variable $a$ as:
\begin{equation}
    2 q_y + a = q_{x_1} + q_{x_2} + q_{x_3},
\end{equation}
leading to linear penalty
\begin{equation}
    P_{\text{spin}}(q_{x_1}, q_{x_2}, q_{x_3}, q_y, a)
    = - a - 2 q_y + q_{x_1} + q_{x_2} + q_{x_3},
\end{equation}
which vanishes exactly when the equality holds.

This construction generalizes to $(2^N - 1)$ inputs using auxiliary variable $a_i$
\begin{equation}
    \begin{split}
        P_{\text{spin}}(q_1, \dots, q_{2^N - 1}, q_y, a_0, \dots, a_{N-2})\\
        = -\sum_{i=0}^{N-2} 2^i a_i - 2^{N-1} q_y + \sum_{j=1}^{2^N - 1} q_j,
    \end{split}
\end{equation}
which evaluates to zero if and only if the encoded sum is consistent with the output bit $q_y$.

\subsubsection{From QCBO Constraints to a Global QUBO Objective}
\label{Sec:QCBOtoQUBO}
Once local penalties are defined, they are combined into a single global QUBO objective. Let $\mat{x}$ denote the set of all Boolean variables, including auxiliary ones:
\begin{equation}
    \label{miniset}
    \mat{x} = \{\tau_0, \dots, q_{z_{0,0}^0}, \dots, q_{r_{\bar{\mat{y}}}}, \dots, a_{0,0}^{0}, \dots\}.
\end{equation}

Inequality constraints in the QCBO formulation (e.g., perturbation budgets) are first rewritten as equality constraints via auxiliary variables and then treated identically to Boolean gate and sign constraints. Collecting all constraints $P_k(\mat{x}) = 0$, we obtain the following:
\begin{equation}
    \label{OCModel}
    \begin{split}
    \underset{\mat{x}}{\text{minimize}} \quad & H_0(\mat{x}) \\
    \text{subject to} \quad & P_k(\mat{x}) = 0, \quad k = 1, 2, \dots, m,
    \end{split}
\end{equation}
where $H_0(\mat{x})=d(q_{\mat{x}}, q_{\mat{x'}})$, and $P_k(\mat{x})$ encode all Boolean, sign, and perturbation constraints.. This constrained problem is converted to an unconstrained QUBO by adding squared penalties:
\begin{equation}
    \label{eq:denseQUBO}
    H(\mat{x}) = H_0(\mat{x}) + \sum_{k=1}^m \lambda_k P_k^2(\mat{x}),
\end{equation}
where $\lambda_k > 0$ balance constraints and the original objective.

The primary objective $H_0(\mat{x})$ is upper bounded by the perturbation budget $\epsilon$ and the number of perturbable pixels. For all instances considered,
\[
0 \leq H_0(\mat{x}) \leq \epsilon \leq P_{\max},
\]
where $P_{\max}$ is the number of pixels allowed to change.

Each constraint $P_k(\mat{x})=0$ is encoded so that violations incur $P_k^2(\mat{x}) \ge 1$. With $\lambda_k=1$, each violated constraint contributes at least one unit of energy, comparable to variations in $H_0(\mat{x})$. Empirically, lowest-energy solutions satisfy all constraints, indicating uniform penalty weights suffice at the studied scales. More generally, $\lambda_k$ can prioritize hard constraints over softer ones; adaptive or constraint-specific scaling for larger problems is left for future work.

\begin{table*}[t]
\begin{center}
\begin{tabular}{|c|c|c|c|c|}
\hline
 \textbf{BNN Layers} & \textbf{Input Image Size} & \textbf{Total Train Data} & \textbf{Total Test Data} & \textbf{Accuracy} \\ \hline
 31$\times$7$\times$10     & $5\times5$   & 1{,}380  & 703    & 43\% \\ \hline
 63$\times$7$\times$10     & $7\times7$   & 23{,}995 & 6{,}212 & 41\% \\ \hline
 127$\times$7$\times$10    & $11\times11$ & 55{,}196 & 9{,}562 & 62\% \\ \hline
 1023$\times$7$\times$10   & $28\times28$ & 59{,}986 & 10{,}000 & 66\% \\ \hline
\end{tabular}
\caption{Configurations of ten-class BNNs used in our robustness experiments. These models are intentionally compact and serve as tractable test networks for the proposed QUBO-based verification pipeline rather than as state-of-the-art classifiers.}
\label{Tab:BNNs}
\end{center}
\end{table*}

\begin{table*}[t]
\centering
\begin{tabular}{|c|c|c|c|c|}
\hline
 \textbf{BNN} & \textbf{Perturbed Pixels} & \textbf{Perturbation Bound} & \textbf{Total Variables} & \textbf{Total Constraints} 
\\ \hline
 $31 \times 7 \times 10$   & $16$  & $8$   & $276$   & $533$    \\ \hline
 $63 \times 7 \times 10$   & $32$  & $32$  & $413$   & $2{,}643$\\ \hline
 $127 \times 7 \times 10$  & $64$  & $32$  & $676$   & $8{,}020$\\ \hline
 $1023 \times 7 \times 10$ & $256$ & $128$ & $2{,}235$& $1{,}027{,}318$\\ \hline
\end{tabular}
\caption{QUBO instances derived from the ten-class BNNs. For each architecture (hidden units $\times$ hidden layer width $\times$ output classes), we report the number of input pixels allowed to be perturbed, the perturbation bound, and the total numbers of Boolean variables and equality constraints in the final QUBO formulation.}
\label{Tab:BNN_Summary}
\end{table*}
\begin{table*}[t]
\begin{center}
\begin{tabular}{|c|c|c|c|c|}
\hline
\textbf{Input Image Size} & \textbf{Variables} & \textbf{Total Constraints} & \textbf{Solver} & \textbf{Constraints Satisfied}  \\ \hline

\multirow{3}{*}{$5\times5$} &
\multirow{3}{*}{276} &
\multirow{3}{*}{533} &
Gurobi & 529    \\ \cline{4-5}
 &  &  & SA    & 533    \\ \cline{4-5}
 &  &  & FEM   & 533    \\ \hline

\multirow{3}{*}{$7\times7$} &
\multirow{3}{*}{413} &
\multirow{3}{*}{2{,}643}&
Gurobi & 2{,}635\\ \cline{4-5}
 &  &  & SA    & 2{,}643\\ \cline{4-5}
 &  &  & FEM   & 2{,}643\\ \hline

\multirow{3}{*}{$11\times11$} &
\multirow{3}{*}{676} &
\multirow{3}{*}{8{,}020}&
Gurobi & 8{,}018\\ \cline{4-5}
 &  &  & SA    & 8{,}019\\ \cline{4-5}
 &  &  & FEM   & 8{,}020\\ \hline

\multirow{3}{*}{$28\times28$} &
\multirow{3}{*}{2{,}235}&
\multirow{3}{*}{1{,}027{,}318}&
Gurobi & 1{,}027{,}316\\ \cline{4-5}
 &  &  & SA    & 1{,}027{,}316\\ \cline{4-5}
 &  &  & FEM   & 1{,}027{,}318\\ \hline

\end{tabular}
\caption{Performance of algorithmic solvers on BNN robustness verification QUBOs. For each input image size, we report the number of QUBO variables, the total number of constraints, and the number of constraints satisfied by Gurobi, Simulated Annealing (SA), and the Free Energy Machine (FEM). FEM consistently finds fully satisfying solutions for all problem sizes, while Gurobi and SA occasionally leave a small number of constraints unsatisfied.}
\label{Tab:AlgoBenchmark}
\end{center}
\end{table*}

\section{Results}
\label{Sec:Results}

Once BNN verification is converted to a Hamiltonian $H(\mat{x})$ with dense couplings as in~\eqref{eq:denseQUBO}, any Ising/QUBO optimizer can, in principle, be applied. In this section, we firstly validate our resulting QUBO model via algorithmic solvers on classical CPU/GPU platforms. Afterwards, we show that the produced models can be deployed on quantum-inspired computing platforms such as D-Wave and Fujistu's digital annealer (DA).

\subsection{Algorithmic Validation on Classical Computers}
\label{Sec:AlgoSolBench}

In this subsection, we validate our quantum-inspired BNN verification frameworks on classical computing platforms. 

{\bf BNN Benchmarks.} Given the combinatorial complexity of the resulting QUBOs and practical solver limits (e.g., Simulated Annealing and Gurobi), we focus on relatively small BNNs to validate the proposed verification pipeline. Although these models are not state of the art, they demonstrate that our QUBO formulation captures BNN robustness properties and is solvable in practice. All BNNs are trained on a binarized MNIST dataset, where grayscale images are thresholded to ${0,1}$ and padded with zeros to form a one-dimensional input of size $2^n - 1$. This design simplifies implementation of the sign function and ensures uniformity across layers. An optional preprocessing step removes contradicting input–output pairs, improving accuracy for smaller models. Table~\ref{Tab:BNNs} summarizes the BNN parameters used in our experiments, including input size, dataset size, and classification accuracy. Architectural choices primarily control QUBO size—driven by input dimension and network depth—rather than maximizing accuracy. Increasing the number of perturbed pixels or the perturbation bound enlarges the feasible space, generally making adversarial solutions easier to find. Table~\ref{Tab:BNN_Summary} reports the number of perturbable pixels, perturbation bounds, and resulting QUBO sizes. Perturbable pixels are selected based on minimal average activation.

 {\bf Solvers.} We solve the resulting QUBO via the following algorithms:
\begin{itemize}
\item \textbf{Gurobi}~\cite{gurobi} is the de facto industry standard for related mixed-integer and QUBO problems, and we use it as the reference baseline. To control runtime, we apply early termination based on objective stagnation, yielding high-quality but not always provably optimal solutions for larger instances.
\item \textbf{Simulated Annealing (SA)} is a probabilistic optimization algorithm that mimics physical cooling processes by gradually lowering a temperature parameter to escape local minima and converge toward low-energy solutions of complex objective functions. In this study, we use D-Wave’s SA with longer schedules and more steps to assess the intrinsic QUBO solvability.
\item \textbf{Free Energy Machine (FEM)}~\cite{Shen2025} is a physics-inspired heuristic for dense QUBOs, combined with hyperparameter search over learning rates and temperature schedules; its core dynamics are summarized in Appendix~\ref{App:FEM}.
\end{itemize}
We search for a minimum-energy solution that satisfies all constraints. A solution is considered \emph{feasible} if all encoded constraints are satisfied, indicating that a valid adversarial perturbation exists and the given BNN is {\it not robust}.

{\bf Algorithm Results.} Table~\ref{Tab:AlgoBenchmark} summarizes the solver outcomes for all QUBO instances. Clearly, FEM and simulated annealing perform better than Gurobi in all BNN verification instances, as their solutions produced satisfy more constraints.  FEM performs best overall: it is able to find fully constraint-satisfying solutions for all QUBO instances. For each such solution, we perform a direct {\bf reverse check} on the original BNN and confirm that (i) the perturbation stays within the prescribed budget and (ii) the predicted label changes, thereby certifying non-robustness of the chosen input under our perturbation model. SA, under the stronger settings used here, also reaches fully satisfying solutions for the $5\times 5$ and $7\times 7$ cases and comes very close for the larger instances. Gurobi returns high-quality but slightly infeasible solutions for all problem sizes, likely due to early termination. Overall, these results confirm that the resulting QUBO problem can be solved to correctly verify the robustness of a BNN.

\begin{figure}[t]
    \centering
    \subfloat[Original input\label{subfig:input}]{
        \includegraphics[width=0.4\linewidth]{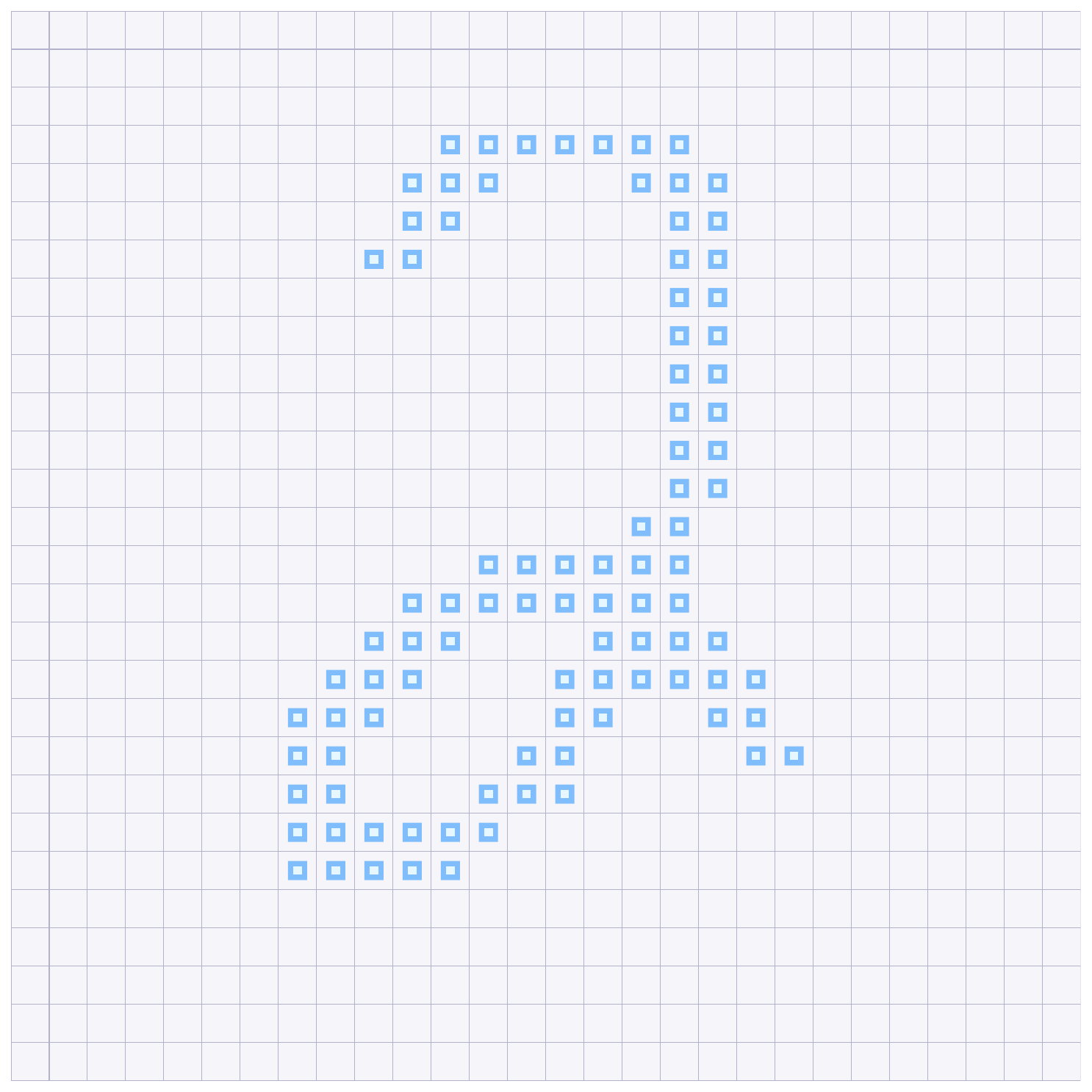}
    }
    \hspace{1cm}
    \subfloat[Perturbed input\label{subfig:perturb}]{
        \includegraphics[width=0.4\linewidth]{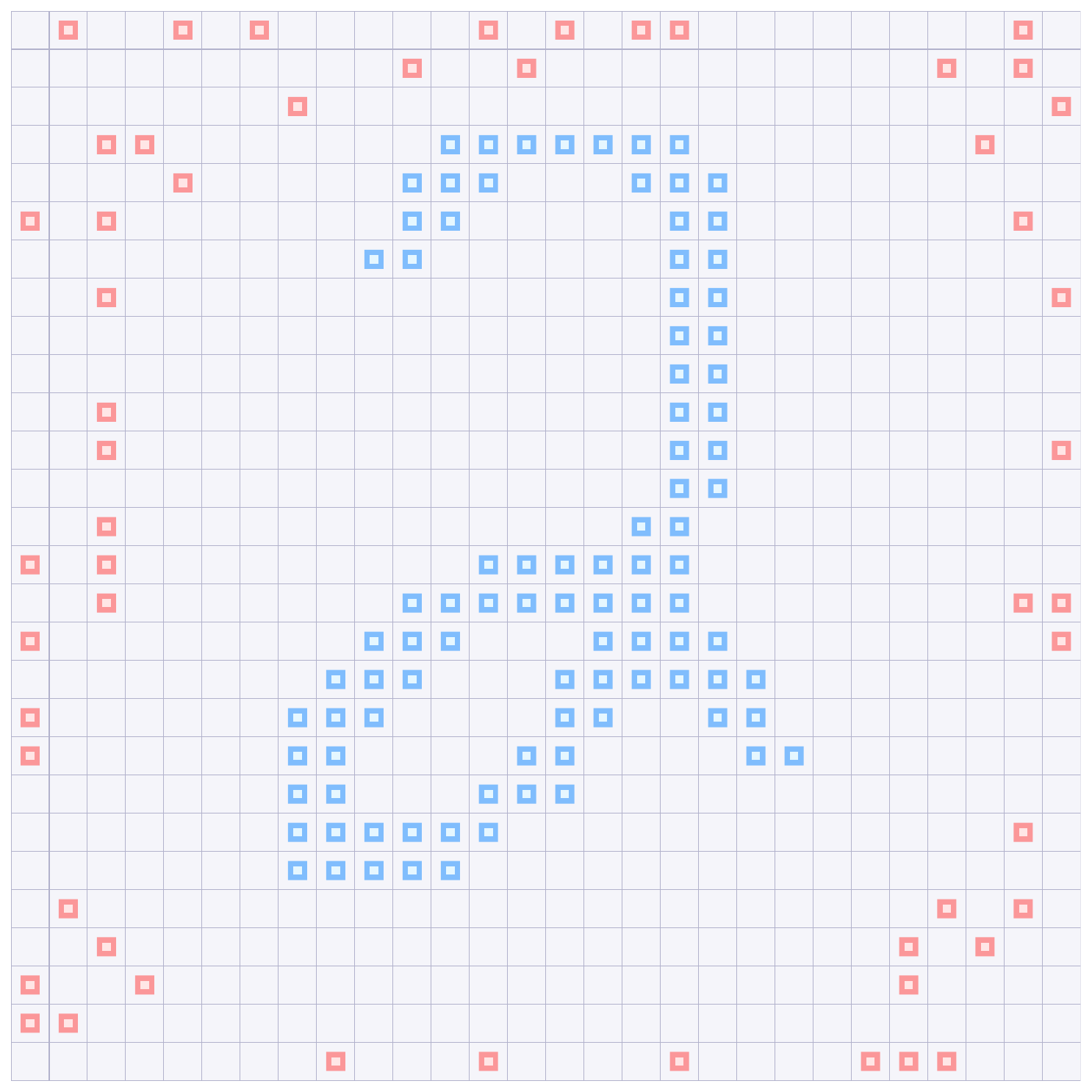}
    }
    \caption{Original (a) input image correctly classified as digit 2 and perturbed (b) input image misclassified as digit 6. Each box corresponds to one input spin: gray background denotes $-1$, blue boxes denote $+1$, and red boxes highlight the perturbations obtained by solving the proposed QUBO with the Free Energy Machine.}
    \label{fig:FullSizeBen}
\end{figure}

\begin{figure}[t]
    \includegraphics[width=\linewidth]{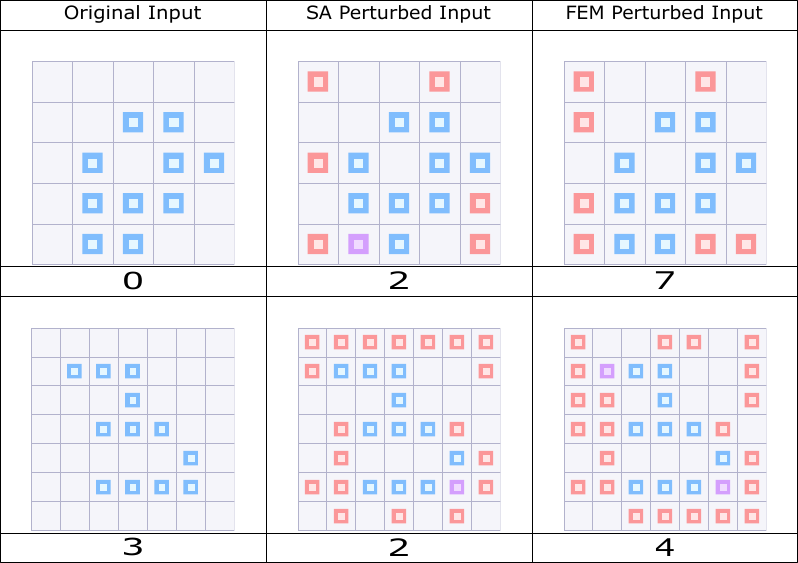}
    \caption{Original and perturbed binarized MNIST inputs, with predicted labels shown underneath each image. The two rows correspond to $5\times 5$ and $7\times 7$ inputs that are initially classified correctly but become misclassified after perturbation. Blue boxes represent pixels with value $+1$, while the gray background represents $-1$. In the perturbed images, red boxes indicate pixels flipped from $-1$ to $+1$, and violet boxes indicate pixels flipped from $+1$ to $-1$. The second and third columns show perturbed inputs obtained using Simulated Annealing (SA) and the Free Energy Machine (FEM), respectively.}
    \label{fig:AlgoBen}
\end{figure}

Since multiple adversarial examples can exist for a non-robust BNN, the solutions produced by SA and FEM are not necessarily identical. As shown in Fig.~\ref{fig:AlgoBen}, the two solvers return distinct perturbation patterns that each lead to a different misclassification of the original input. In these small-size cases, the number of perturbed pixels is still relatively large compared to the structure present in the input image,  so the existence of multiple low-energy minima is expected and the corresponding QUBOs are comparatively less challenging than the full-size case. For the full-size MNIST input BNN, Fig.~\ref{fig:FullSizeBen} shows a sparse perturbation suffices to alter the BNN’s prediction, providing a more compelling and realistic demonstration of robustness failure for high-dimensional inputs.

\subsection{Deployment on Quantum/ Quantum-Inspired Hardware}
\label{Sec:SimHardBench}
\begin{table*}[t]
\begin{center}
\begin{tabular}{|c|c|c|}
\hline
 \textbf{Solver} & \textbf{Constraints Satisfied} & \textbf{Total Time (Seconds)} \\ \hline
Gurobi                &  1{,}273    & 61.447 \\ \hline
Quantum Annealer      &  356$^*$  & 0.724$^\dagger$ \\ \hline
Digital Annealer      &  1{,}273    & 0.366          \\ \hline
Simulated Annealing   &  1{,}273    & 3.559          \\ \hline
\end{tabular}
\caption{Comparison of hardware executions and classical baselines for BNN robustness verification on a representative QUBO instance. Gurobi is included as an industry-standard mixed-integer/QUBO optimizer, while Simulated Annealing (SA) serves as the annealing-style baseline. For each solver we report the number of constraints satisfied by the best returned sample and the total runtime. $^*$Best sample among all samples obtained from the quantum annealer. $^\dagger$Total runtime accumulated over 5{,}000 shots.}
\label{Tab:HardwareResults}
\end{center}
\end{table*}

\begin{figure}[t]
    \centering
    \subfloat[Logical QUBO\label{subfig:qubo_logical}]{
        \includegraphics[width=0.4\linewidth]{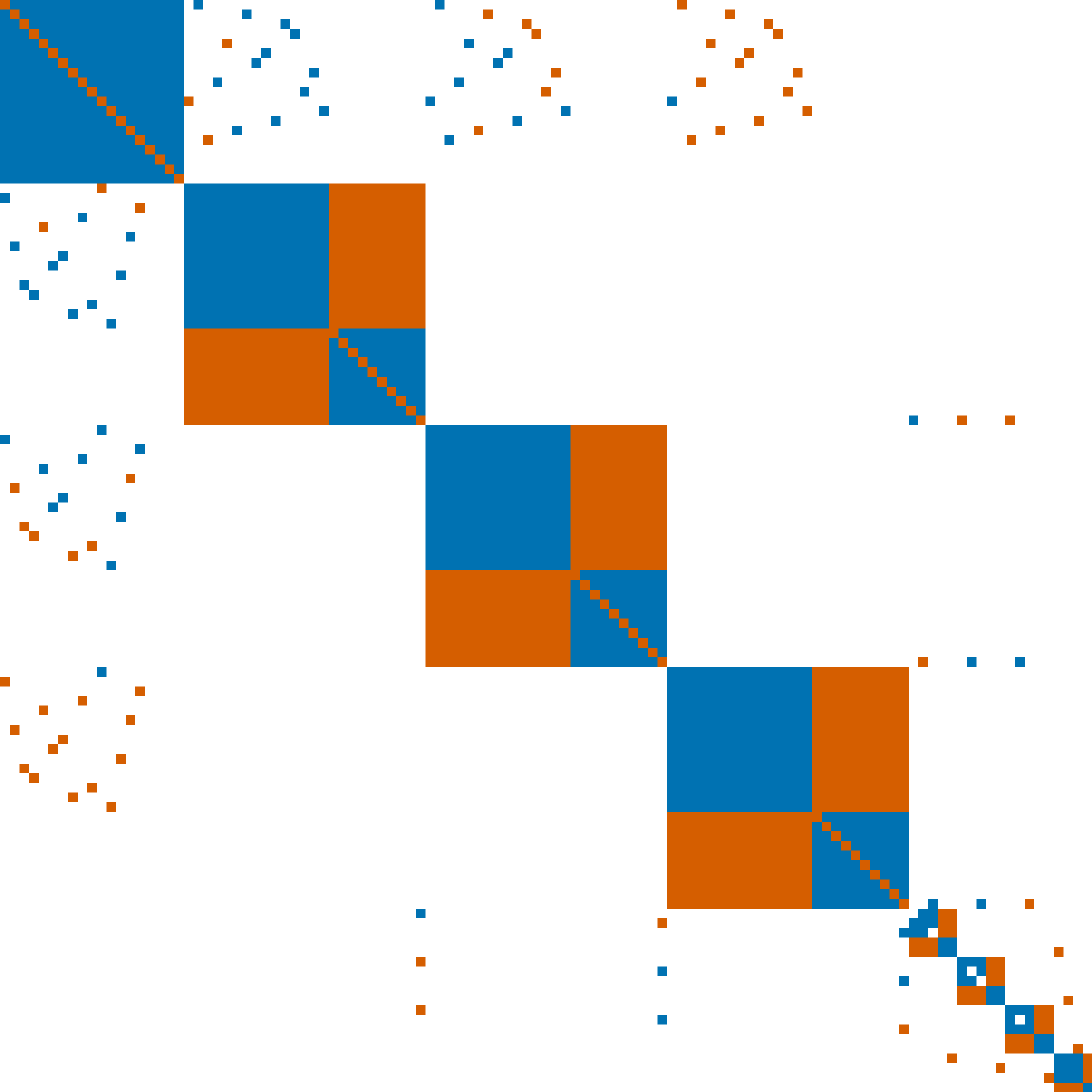}
    }
    \hspace{1cm}
    \subfloat[Embedded QUBO\label{subfig:qubo_embedded}]{
        \includegraphics[width=0.4\linewidth]{Figures/embedded_qubo.pdf}
    }
    \caption{Logical and embedded QUBO instances for BNN verification. (a) Logical QUBO coupling matrix, where blue and orange entries indicate positive and negative couplings, respectively. (b) Embedded QUBO on a subset of D-Wave’s Pegasus topology; nodes denote physical qubits and edges denote physical couplers realizing the logical 
    interactions.}
    \label{fig:QUBO_Embedding}
\end{figure}

{\bf BNN Benchmarks.} In principle, our produced QUBO can also be solved with various quantum-inspired hardware. To demonstrate this,  we consider a simpler BNN instance: a two-class BNN (namely with output labels ``0" and ``1") with full size MNIST input ($28 \times 28$), three hidden units and three hidden layers. The perturbation budget allows up to 15 input bits to flip, resulting in a QUBO with 113 variables and 1{,}273 equality constraints.

{\bf Hardware Solvers.} We deploy the resulting QUBOs on two popular annealing hardware:
\begin{itemize}
\item \textbf{D-Wave Quantum Annealer} solves the resulting QUBO by evolving a quantum system from an initial transverse-field Hamiltonian toward the problem Hamiltonian via quantum annealing. During this process, quantum superposition and tunneling help the system explore the energy landscape and settle into low-energy states corresponding to candidate solutions.
\item \textbf{Fujitsu’s Digital Annealer} is a classical, CMOS-based system that emulates the dynamics of annealing by directly searching the energy landscape of QUBO/Ising formulations using massively parallel, high-speed updates. It combines deterministic and stochastic update rules to efficiently escape local minima and find near-optimal solutions without relying on quantum effects. 
\end{itemize}
Since the D-wave hardware has a sparse connection, the resulting highly coupled QUBO from BNN verification needs to be embedded into a sparse one, resulting in increases in variables and non-zero couples (c.f. Fig.~\ref{fig:QUBO_Embedding}). This is why we limit our experiments to smaller-size BNNs in this subsection. 

\begin{figure}[t]
    \centering
    \subfloat[Original Input\label{input}]{
        \includegraphics[width=0.4\linewidth]{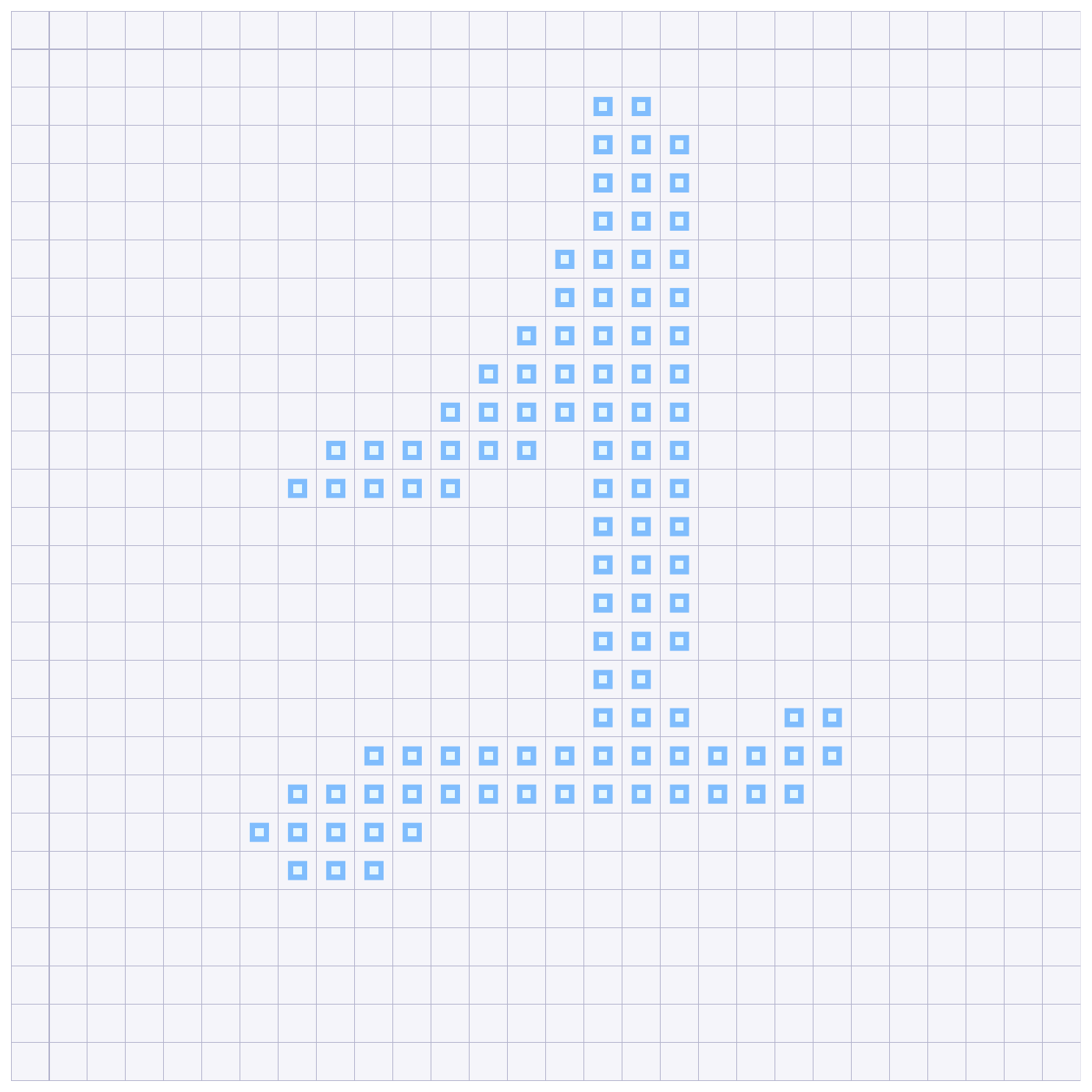}
    }
    \hspace{1cm}
    \subfloat[Perturbed Input\label{perturb}]{
        \includegraphics[width=0.4\linewidth]{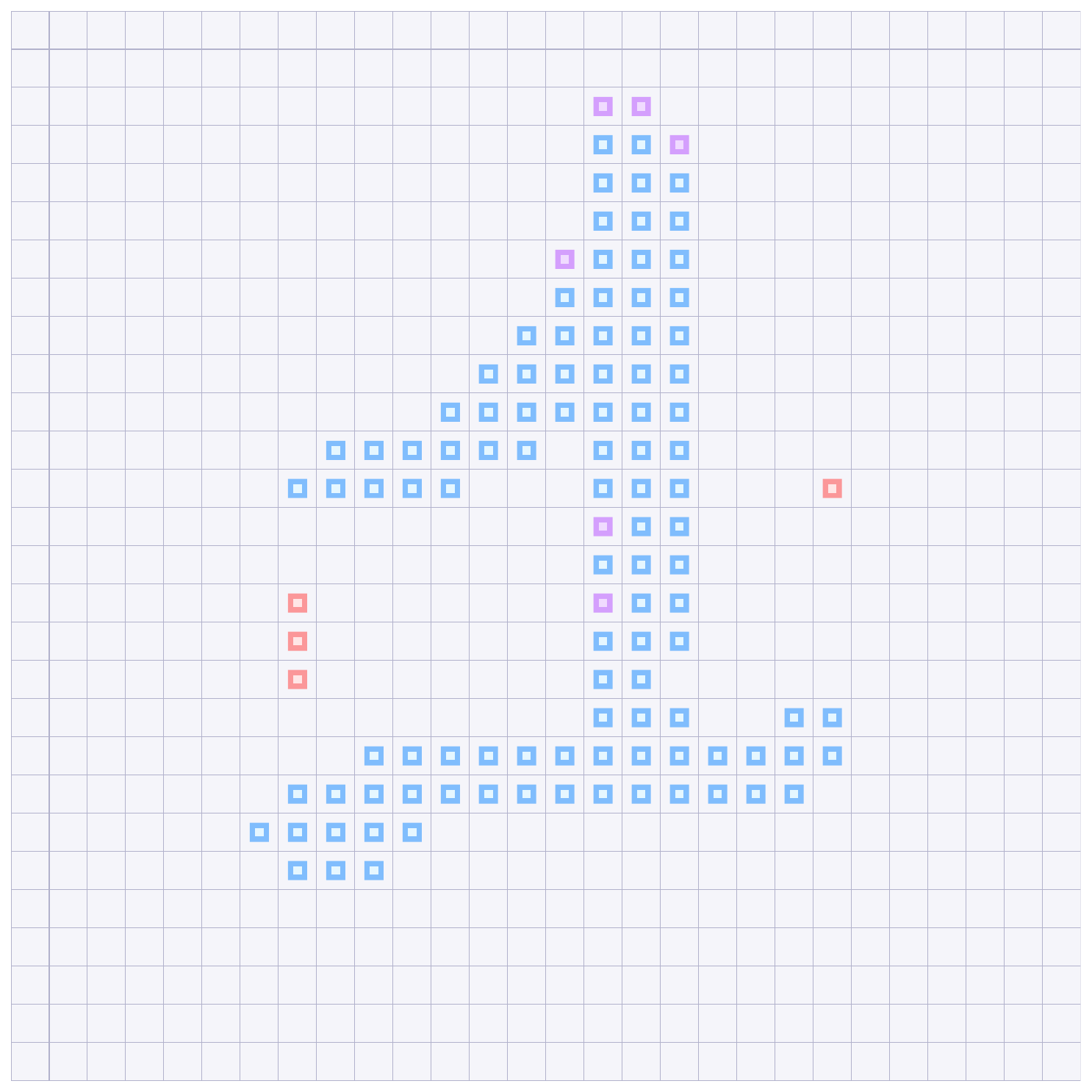}
    }
    \caption{Perturbation obtained from solving the simplified QUBO instance via digital annealing.}
    \label{Fig:DASol}
\end{figure}

{\bf Results.} Table~\ref{Tab:HardwareResults} summarizes the outcomes obtained on this instance using the quantum annealer, the digital annealer, Gurobi, and Simulated Annealing (SA). The reported ``Total time'' reflects the wall-clock execution time on the respective hardware or software platform. The digital annealer finds a fully constraint-satisfying solution (see Fig.~\ref{Fig:DASol}) and, on this instance, attains a wall-clock speedup of roughly $168\times$ over Gurobi running on a 64-core Intel Xeon server, while also being about $10\times$ faster than our SA baseline, all at comparable solution quality (all constraints satisfied). These results indicate that dedicated quantum-inspired hardware can efficiently handle the QUBO instances arising from our BNN robustness formulation, at least at this scale. The quantum annealer returns its best sample in $\sim 20~\mu\mathrm{s}$ per shot, illustrating the very low per-sample latency of current quantum hardware. However, the best quantum sample still violates some constraints. The combination of fast sampling and improved qubit performance suggests that future generations of quantum annealers may become competitive solvers for QUBO-based robustness verification.

In general, the results in Table~\ref{Tab:HardwareResults} indicate that our QUBO encoding is compatible with Ising and quantum-inspired hardware. However, hardware accelerators with sparse connections (e.g., D-wave quantum annealer and many existing sparse ising machines) are less efficient due to the dense connections in our generated QUBO instances. 

\section{Conclusion}
In this paper, we have proposed a QUBO-based formulation to verify the robustness of binary neural networks (BNNs). The resulting QUBO instances provide a unified representation compatible with Ising machines and quantum-inspired computing platforms. Using classical baselines (Gurobi and Simulated Annealing) and the free energy machine (FEM), we have shown that these QUBOs are practically solvable and that computing results can correctly tell if the BNN is robust or not.

This work has established a concrete synergy between AI trustworthiness and emerging quantum/ quantum-inspired computing by mapping BNN robustness verification directly to QUBO and deploying the resulting instances on both a D-Wave quantum annealer and Fujitsu’s Digital Annealer. We have observed that the fully connected Fujistu DA offers better accuracy and efficiency than the sparse connected D-wave quantum annealer in our problems. We expect that future architectures with improved connectivity, reduced noise, and more efficient embedding strategies will narrow the gap to the algorithmic performance achieved by FEM and related software solvers, further strengthening the role of annealing hardware in neural network robustness verification.

\section*{Data availability}
The MNIST dataset used to train the BNNs in this work is publicly available from the MNIST website~\cite{MNSIT_2012}. All code required to construct the BNN robustness verification QUBOs and reproduce the experiments is available at: \url{https://github.com/Rahps97/BNN-Robustness-Verification.git}.

Due to size limitations, all the generated QUBO instances are not included in the repository; however, they can be recreated directly using the provided scripts. Exact copies of the QUBO instances used in this study are available from the authors upon reasonable request.

\section*{Author Contribution}
R.S. conceived the detailed idea of the BNN robustness-verification-to-QUBO formulation, implemented the end-to-end QUBO pipeline, and carried out all numerical experiments. S.S. developed the initial two-class BNN verification formulation. Z.Z. initiated the project, conceived the high-level idea, raised the funding, supervised the project, contributed to the problem formulation and interpretation of the results, and provided critical revisions to the manuscript. All authors discussed the results and approved the final version of the manuscript. S.S. conducted this work while affiliated with the University of California, Santa Barbara. 

\section*{Acknowledgment}
The authors sincerely thank Dr. Zi-Song Shen for valuable discussions on the Free Energy Machine. We are also grateful to Dr. Hayato Ushijima and Dr. Indradeep Ghosh of Fujitsu for executing our experiments on the Digital Annealer, and to Dr. Hirotaka Tamura (DXR Laboratory Inc.) for facilitating the collaboration with Fujitsu Research Labs. This work was supported by NSF Grant 2311295 and DOE Grant DE-SC0021323.

\section*{Competing interests}
The author declare no competing interests.

\appendices
\section{Free Energy Machine Dynamics}
\label{App:FEM}

\begin{figure}[t]
\centering
\begin{femalgorithm}{Free Energy Machine}{FEMCore}
$J,h, T_{\mathrm{init}},T_{\mathrm{final}}, N_{\mathrm{step}}, \eta,\mu,\lambda,c_{\mathrm{grad}},\sigma \longrightarrow s^\star,E(s^\star)$

\BlankLine
$\ell\sim\mathcal{N}(0,\sigma I),\quad m\leftarrow 0,\quad v\leftarrow 0$,\\
$\beta_t^{-1}\leftarrow (1-\alpha_t)T_{\mathrm{final}}+\alpha_t T_{\mathrm{init}}, \quad \alpha_t=\tfrac{t}{N_{\mathrm{step}}-1}$

\For{$t=0,\dots,N_{\mathrm{step}}-1$}{
  $p\leftarrow \sigma(\ell),\quad s\leftarrow \mathrm{round}(p),\quad u\leftarrow Js$\\
  $g \leftarrow -c_{\mathrm{grad}}\left(2u+h+\tfrac{\ell}{\beta_t}\right)\odot p\odot(1-p)$\\
  $v \leftarrow \alpha_t v + (1-\alpha_t) g^{\odot 2}, \quad \ell \leftarrow \ell - \eta\, g/(\sqrt{v}+10^{-8})$\\
  $\ell \leftarrow (1-\lambda)\ell + \mu m, \quad m\leftarrow \ell$
}
\BlankLine
$s^\star\leftarrow \mathrm{round}(\sigma(\ell)),\quad E(s^\star)\leftarrow -h^\top s^\star - \tfrac12 (s^\star)^\top J s^\star$
\end{femalgorithm}
\end{figure}

For completeness, we provide here the pseudocode of the core Free Energy Machine (FEM) solver used in our experiments (see Sections~\ref{Sec:Results} and~\ref{Sec:AlgoSolBench}). The algorithm, summarized in Algorithm~\ref{alg:FEMCore}, implements a stochastic annealing dynamics on continuous logits $\ell_i$, coupled through the Ising couplings $J$ and biases $h$. At each step, FEM updates the logits using a pseudo–free-energy gradient, an inverse-temperature schedule, and an RMSProp–style adaptive step with momentum and weight decay, and finally rounds the resulting probabilities to obtain a binary spin configuration and its corresponding energy. This explicit description is intended to clarify the implementation details underlying our FEM-based robustness verification results.

\bibliographystyle{ieeetr}
\bibliography{cite}

\end{document}